# The Great Deception: A Comprehensive Study of Execution Strategies in Corporate Share Buy-Backs

## Decoding the Enigma of Equity Repurchases and their Disproportionate Cost Structures

Michael Seigne[1], Joerg Osterrieder[2]

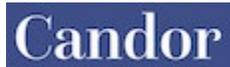

June 30, 2023

---

[1] Candor Partners Limited, United Kingdom
[2] University of Twente, Netherlands and Bern Business School, Switzerland

# 1. Introduction

Share buy-backs have gained significant prominence as a strategic tool in the corporate capital allocation landscape over the past few decades. Despite extensive research and debates on their perceived merits and shortcomings in financial journals, investor letters, and the media, surprisingly little attention has been given to the actual execution of these transactions, now compare that to the volumes on other capital allocation decisions such as acquisition pricing. In most jurisdictions, there is a lack of transparency regarding the trading strategies and products employed to implement these important decisions, in contrast to all the attention they attract. We simply ask the questions: Why this lack of research on the execution phase and what can we learn?

This white paper takes a unique perspective as it is authored by outsiders looking into the world of share buy-backs. Although we lack firsthand experience in executing share buy-backs, we possess extensive expertise in equity execution, particularly in quantitative and strategic aspects of execution quality and analytics. From the outset of this work, we identified a discrepancy in the execution practices of certain share buy-backs, and our aim is to shed light on this issue.

The purpose of this paper is to assist corporations, investors, and regulators in understanding the "dark arts" of execution when it comes to share buy-backs. Corporations are the backbone of our capital markets, providing the necessary investment opportunities and the very securities we all trade. It is imperative that we develop a service that enables corporations to trade their own listed securities in a manner that is both regulatory compliant and cost-efficient, similar to the low-friction experience enjoyed by other market participants. Why is it that retail investors are charged zero fees, institutional investors pay single-digit basis points, yet corporations often incur fees that are five, ten, or even a hundred times higher for similar or inferior outcomes? This seems upside down right? The landlord pays the rent while the tenants live for free.

In the subsequent sections of this paper, we will delve into the intricacies of share buy-back executions, highlighting some of the challenges and opportunities that arise. By analysing current practices and some of the issues, we aim to foster a better understanding of the execution phase of corporate share buy-backs and advocate for fair and transparent processes that benefit all stakeholders in the capital markets ecosystem.

## 1.1 The Scale of the Buy-Back Market

It is with good reason that share buy-backs continue to receive a significant amount of stakeholder focus as they grow in importance within corporate's strategic capital allocation decisions. The sheer size of financial resources committed to buy-backs underscores their growing importance. In the United States alone, S&P 500 companies have spent $3.8 trillion dollars over the past 5 years[3] to repurchase their own shares.

However, the significance of share buy-backs extends beyond their size. When properly executed, buy-backs can enhance earnings per share, potentially boosting stock prices, and providing an efficient mechanism for companies to return excess cash to shareholders. Conversely, improperly executed buy-backs can dilute long-term shareholder value, and erode capital, undermining the very purpose of this strategy.

## 1.2 Regulatory Focus

Alongside the substantial growth of share buy-backs, recent years have seen a significant surge in regulatory attention and political pressure. This increased focus is fuelled not only by the rising prominence of buy-backs but also by evolving environmental, social, and governance (ESG) considerations. The shifting landscape has resulted in greater scrutiny from shareholders, more press commentary, regulatory adjustments, and tax changes, all aimed at ensuring transparency, fairness, and accountability during the share buy-back process.

---

[3] Yardeni, E. (2023). Corporate Finance Briefing: S&P 500 Buy-backs & Dividends. Yardeni Research, Inc. https://www.yardeni.com/pub/buy-backdiv.pdf

As an example, in May 2023, the U.S. Securities and Exchange Commission (SEC) enacted new transparency rules[4] specifically pertaining to share buy-backs. This move reflects the regulatory body's commitment to enhancing corporate accountability and market transparency.

Meanwhile, the Australian Securities and Investments Commission (ASIC) has demonstrated a similar focus, with five of its last eight fines related to the execution of share buy-backs. This highlights the regulator's vigilance in monitoring buy-back practices and its readiness to penalise breaches.

In the UK, the Financial Conduct Authority (FCA) is rolling out its "Improving Equity Secondary Markets"[5] initiative. This program underscores the FCA's aim to create a more robust, fair, and effective market, which inevitably includes the processes surrounding share buy-backs.

## 2. The Capital Allocation Decision in the context of share-buy-backs

Section 2 delves deeper into the intricacies of the Capital Allocation Decision, particularly in the context of share buy-backs. We explore the role of Board and Corporate Executives (the Company) in this complex process, their duty of diligent oversight, and the importance of their strategic vision. Additionally, we discuss how the decision-making must extend throughout the implementation phase to ensure long-term shareholder value maximisation. This section also covers critical parameters influencing the buy-back decision and the need for an integrated approach to risk minimisation and value creation. However, before we address these parameters we quickly need to account for the various differences of opinion regarding the merits of share buy-backs in the first place. As we have already mentioned, we do not want to join the debate[6] regarding this topic, merely find a way to categorise those views, so we can then discuss the execution strategy of share buy-backs through these various different lenses.

### 2.1 Three Main Camps

We broadly divide the opinion into three camps[7].

1. The "Value" Camp- argue that share buy-backs create value for the long term shareholder if they are carried out when the market capitalisation of the company is trading cheap relative to "Intrinsic Value". Warren Buffet[8] and Lord Wolfson of Next Plc[9] have at times argued the core views of this camp. [10]
2. The "Excess Capital" Camp- argue that share buy-backs are a more tax efficient and flexible way to return excess capital to shareholders than dividends or special dividends. Shareholder value is not created in the act of buying back shares, rather the value of the excess capital merely passes from the corporation back to shareholders, who can choose to either retain their increased ownership, or sell to create a "synthetic" dividend. There is a "wealth" transfer between the remaining shareholders and those that sell, depending on the transaction price of the buy-back relative to the "intrinsic value".

---

[4] Securities and Exchange Commission. (2023). Share Repurchase Disclosure Modernization. 17 CFR Parts 229, 232, 240, 249, and 274. Release Nos. 34-97424; IC-34906; File No. S7-21-21, RIN 3235-AM94. https://www.sec.gov/rules/final/2023/34-97424.pdf

[5] https://www.fca.org.uk/publications/policy-statements/ps23-4-improving-equity-secondary-markets

[6] CFA Institute - Stock Buy-back Motivations and Consequence: A Literature Review, Harvard Business Review - The Case for Stock Buy-backs, FT - Do most buy-backs destroy value? FT- Share buy-backs: welcome payments in a suspect currency, Knowledge at Wharton: Making Sense of Stock Buy-backs

[7] Michael Mauboussin and Dan Callahan - section on share buy-backs call the camps "schools" and argue this whole topic better than we ever could. Morgan Stanley Investment Management : Capital Allocation , AQR's Buyback Derangement Syndrome

[8] Berkshire Hathaway: 2011 Letter to Shareholders

[9] Next Plc. (2013). Annual Report and Accounts January 2013 [Annual Report]. Retrieved from https://www.annualreports.com/HostedData/AnnualReportArchive/n/LSE_NXT_2013.pdf

[10] Included in this camp are the other reasons that share the same goal of reducing the share capital. E.g European banks working on improving their CET1 ratios. We should really call this the "Reduce Share Count" Camp, but that is too much of a mouthful.

3.  The "Never" Camp - we are not going to discuss the many objections here, as this paper is focused on the process once the Company has made the decision to implement a buy-back.

## 2.2 Role of Board and Corporate Executives

The Company is entrusted with the responsibility of steering the company in a manner that protects and enhances long-term shareholder value. The decisions revolving around capital allocation directly influence the corporate strategic direction and the creation of long-term shareholder value. One of the decisions that can be made in this process is the buy-back of the company's own shares. When the Company decides on share buy-backs, it indicates their belief that this form of capital allocation would yield the highest return for shareholders, either due to the company's current share price implied valuation, and/or due to the efficient mechanism to transfer capital back to shareholders.

The point we are making here is that the implications of the decision to engage in share buy-backs does not cease at the time of decision to do a share buy-back. Rather, the effects of this decision extend until the conclusion of the execution phase.

## 2.3 Extending the Capital Allocation Decision Throughout the Implementation Phase

Changes in the company's relative valuation, or changes in the implementation costs, during the implementation period can significantly alter the expected return for shareholders. Therefore, the responsibility of the Company also involves diligent oversight of the execution phase to ensure that shareholder value is maximised, reinforcing their commitment to act in the best interests of shareholders.

Defining and considering the critical parameters that shape the buy-back decision is a vital part of planning the execution phase. These parameters need to be driven off what the primary logic of the buy-back decision is. This takes us back to the points we made earlier on the "Value" or the "Excess Capital" camps.

**Value Camp:**

> If the decision is made from the perspective of the "Value" camp then some of the critical parameters are: the target value of capital to be returned, the relative valuation methodology, the frictional costs and the urgency or maximum timeframe for the buy-back execution need to be understood by the implementation team prior to planning the execution strategy. This is because the relative valuation methodology and frictional costs, can be impacted by the inevitable fluctuations in the share price over the course of the execution period.
>
> In the "Value" camp maximising the number of shares purchased for the allocated value of the share buy-back will result in the greatest long-term shareholder value creation. The logic being that the larger the number of shares that are repurchased, and then cancelled, the greater the reduction in the share count, which has the largest beneficial effect on the residual shareholders as their percentage ownership of the company increases.
>
> The principal reason that the valuation methodology needs to be understood by the implementation team is that the share price is likely to fluctuate during the execution period. The execution strategy and broker contracts used to execute the buy-back need to be able to uphold the Company's objectives, if for example the share price was to double. Logically, if the original decision is founded in the "Value" camp one would expect that as the share price rallies the valuation gets less and less attractive, and above some share price the Company would no longer believe that a share buy-back is still in the best interest of shareholders at all. Conversely if the share price was to fall significantly then for valuation reasons the confidence in the decision grows stronger. For a very simple model of this theory working look at our hypothetical example in Appendix 2.

**Excess Capital Camp:**

> In the "Excess Capital" camp maximising the transfer of excess capital to the existing shareholders is the ultimate goal. The logic is that so long as the share price at which the buy-back is transacted at is at a fair market price, then the actual price is not relevant. Buying 1mil shares at a price of £100, has the same capital transfer effect as buying 2mil shares at £50. The actual number of shares is therefore not the primary issue, rather attempting to minimise the frictional loss on the capital's value as it transfers through the process is the goal.
>
> The other main requirement for this logic to hold true is that in order for a shareholder who wants to sell into the buy-back to make this "more efficient" transfer system work, shareholders need to be able to calculate the right number of shares, and the share price of the company's buy-back as it progresses. The manner and timing of this required information that a seller needs is dependent on both the jurisdictional rules of the market and the method[11] used for implementing a share buy-back.
>
> There is always a time lag between when the company completes part or all of its buy-back, and shareholders knowing these details. As we have said this time lag varies by jurisdiction, so for example in the UK you can currently receive enough information on the next business day as the share buy-back progresses. In the US you can not, and it can sometimes be months. Any time delay introduces the concept of share price risk. Later in this paper we will discuss the relationship between share price risk and the length time of the exposure to that risk. Most of that discussion is framed in the context of execution strategies, however this risk is also relevant to the information time lag discussed in this section.

It is clear, when looking at both "Camps" that there are a number of common topics to do with share price risk and frictional costs that the company needs to understand, and the dynamics of how they may change over the course of the implementation phase of a buy-back. The implications of these topics must be weighed against the potential benefits and drawbacks of alternative methods of returning capital to shareholders, such as dividends and special dividends. An integrated and ongoing approach to the decision-making and its implementation helps to minimise risks associated with share buy-backs and maximises the potential for a beneficial outcome.

## 2.4 Risks, Costs and Taxes of Implementing Share Buy-Backs

We aim to provide more insight into the key factors directly related to the buying of shares in the context of share buy-backs, specifically focusing on taxes, costs, and risks. We shall deliberate these factors while excluding discussions around topics such as capital gains tax vs. income tax for shareholders, which have been extensively covered in prior literature. Moreover, the impact cost of trading will be discussed under the risk section to provide a more comprehensive analysis. When we discuss friction, we are referring to taxes, broker fees and or commissions.

**2.4.1 Taxes**

In some jurisdictions such as the US there are now specific taxes on share buy-backs, a new 1% tax became effective in Jan 2023. In the UK there is a 0.5% stamp tax on all buy transactions for shares, regardless of reason for the purchase. There is little one can do to manage this cost, only that it needs to be factored into the total costs while the Company considers the capital allocation decision.

**2.4.2 Understanding the Cost Structure of Share Buy-backs**

Implementing share buy-backs entails a suite of expenses that primarily encompass commissions and fees to compensate brokers and advisors for their services and risk management responsibilities. These costs often pivot on the chosen execution strategy and the details of the broker contract. The costs associated with a share buy-back generally fall into two categories:

- **Flat Commissions or Fees:** This straightforward cost structure is usually determined and understood before the Company's decision-making process. While it's essential to account for these costs, they often represent a predictable part of the financial landscape of a buy-back strategy.
- **Fees derived from trading activity:** These costs are inherently linked to the brokers' trading activities during the execution phase, and demand careful scrutiny. They can constitute a substantial portion of the buy-back value and thus have significant relevance for the overall strategy.

---

[11] This paper only focuses on methods that involve acquiring shares directly or indirectly from the stock market. Specifically, we are not addressing alternative methods such as Dutch Auctions.

Regardless of the type of fees involved, the Company must evaluate their long-term impact on shareholder value. It is crucial to underline that for both the company and its shareholders, the execution phase of share buy-backs are purely cost-centric activities devoid of profit generation. We will elaborate further on this issue, along with topics such as "out performance sharing" fee structures, in later discussions.

### 2.4.3 Price Risk

Navigating the course of a share buy-back operation involves contending with various risks that emerge **during the implementation phase**. The pivotal risk among these is the "price risk," influenced by fluctuations in the share price during the execution of the buy-back.

- **Understanding Price Risk**: The potential for price risk comes from the inherent volatility of the stock market. This volatility plays a significant role in influencing the buy-back's final outcome. As such, it is essential that the Company comprehends, quantifies, and manages this risk effectively. The price risk is driven by two primary factors:
    - **Stock Price Volatility:** The level of risk increases with a rise in the share price's volatility. Higher volatility leads to greater uncertainty and, therefore, greater price risk.
    - **Time Frame Exposed:** The duration of the exposure to risk is another critical element. A longer time frame opens the door for more substantial price swings, thus amplifying the price risk. A greater explanation of how longer and shorter timeframes affect VaR are shown in Appendix 1.

- **Managing and Quantifying Price Risk:** To measure and manage these risks, we employ the Value at Risk (VaR) methodology. This statistical technique, discussed in more detail in Appendix 1, helps estimate the potential downside risk during the execution of the buy-back. An important aspect of risk in this context is its unidirectional nature; it only takes into account the movement of the stock price and not the underlying reasons for such movement.

To give a tangible example of price risk resulting from an execution strategy, let's consider a buy-back contract period of 125[12] days, assuming the stock's annual volatility is 35%[13]. This scenario implies that a one standard deviation move over 125 trading days share price could move by about 25% in either direction, thereby creating a 50% price range. This, in turn, means that the buy-back operation could procure 25% more or less shares in 125 days time. With a typical buy-back size of US$870m[14], the VaR corresponding to a one standard deviation price move at the initiation of the buy-back contract stands at $215m.

---

[12] Average Accelerated Share Repurchase (ASR) Characteristics mean contract period 125 days, Chen, Kai, Press release management around accelerated share repurchases (April 7, 2020). European Accounting Review (2021) 30(1): 197–222, Available at SSRN: https://ssrn.com/abstract=3572543
Kai Chen (2020) Press release management around accelerated share repurchases.

[13] At the time of researching the median volatility of a stock in the S&P 500 and the STOXX 600 was 30.5 and 29.5% respectively. Note this is not the same as VIX, which was 14.3%. VIX is normally much lower as it measures the index as a whole, so the portfolio effect of all constituents dampens the index volatility.

[14] S&P Globals Capital IQ, screen Company Geography: United States, Transaction Type In Buy-back, Buy-back Feature In Accelerated Share Repurchase; Market Repurchase – total count 17,556, filter by 2022 990 count, filter out value N/A 805, average value $870.088mil.

As we have mentioned there is a more detailed explanation of price risk, and how this varies with time frame, especially in the context of an overall execution strategy is discussed more in Appendix 1.

## 2.5 The Impact of Trading

The notion of "impact" in trading has significant implications and depends on the scale of trading activity. The effect of trading, in this context, is analogous to how differently a feather and a meteor hit the ground; both have an impact, but the scale and consequences differ drastically. All trading activities inevitably influence the share price.

Fast and repeated buying of shares can significantly affect the share price, necessitating careful management of this activity. There is substantial research available on this subject (Harvey et al, 2021)[15], which we won't delve into here, but in brief: the larger the daily quantity of shares purchased, the greater the impact on the share price, all other things being equal.

This speed variable in execution is often managed through the "participation rate." This term refers to the proportion of shares executed in a day relative to the total shares traded in the market that day. For instance, a participation rate of 10% of the daily volume is faster, and thus has a greater impact, than a participation rate of 1%. Consequently, the higher the participation rate, the greater the effect on the share price. There are regulatory rules or safe harbours that also come into play regarding participation rate and other factors, see Appendix 3 for more details. The intricate balance between participation rate and impact underscores the complexity of developing an efficient execution strategy.

## 2.6 Trade-off between Impact and Risk

When executing a share buy-back, there's a critical balance that must be maintained: minimising trading impact while mitigating share price risk. A lower participation rate in trading will reduce impact, but the longer the buy-back execution takes, the greater the risk from share price fluctuation. This delicate trade-off between speed of risk reduction and the increased impact from trading faster, coupled with any other company-specific objectives relating to share price value, should be carefully managed within a well-designed share buy-back execution strategy.

## 2.7 Execution Strategy Design and Evaluation of Design

**Execution Strategy Design.** The design of an effective execution strategy for a share buy-back requires careful attention to a range of pivotal factors. An aspect that needs to be accounted for in the strategy design process is the role of execution objectives and incentives common in today's market. These factors, while not immediately apparent, can significantly influence the outcomes for shareholders, we will discuss broker compensation and the relevance of this on some strategies in a later section.

Additionally, some consideration must be given to the effects of impacting the share price due to the execution phase of a buy-back and any subsequent price support to the share price. As we have already discussed, all share trading has an impact on the share price, and in most jurisdictions, it is illegal for a company to provide price support to their shares under normal trading conditions[16]. It is clearly not possible to avoid the fact that buying shares creates some sort of price support, however specifically designing a buy-back strategy with the explicit purpose of providing price support could potentially lead to regulatory issues and be seen as market abuse, and so should be avoided.

---

[15] Harvey, Campbell R. and Ledford, Anthony and Sciulli, Emidio and Ustinov, Philipp and Zohren, Stefan, Quantifying Long-Term Market Impact (September 21, 2021). Available at SSRN: https://ssrn.com/abstract=3874261 or http://dx.doi.org/10.2139/ssrn.3874261

[16] The general exception to this rule in most jurisdictions are "Green-Shoe" or "Over Allocation" structures sometimes used when primary shares are issued and sold.

For long-term shareholders in the "Value" Camp, their interests post the announcement of a share buy-back are pretty straightforward – buy as many shares as possible. Prolonged execution times could negatively impact these interests in two ways. Firstly, the quicker the shares are bought back and cancelled, the sooner their percentage ownership in the company increases. Secondly, any delay in execution time frame extends the period during which shareholders are exposed to potential share price fluctuations. Nobody, including shareholders, wants to bear share price risk without an expected reward. If the company doesn't have a view that the direction of the share price path will be lower within the expected window of the buy-back execution, then the most responsible strategy is to reduce the share price risk as quickly and efficiently as possible. Conversely if the company truly thinks that the share price is trading at a significant and expected to be short lived discounted share price then the ability to accelerate the programme should be factored into the strategy if possible.

**Post-Trade Evaluation.** Once the execution phase is complete, a comprehensive post-trade evaluation is essential to evaluate the outcome. The different objectives of the "Value" and "Excess Capital" camps should be considered in slightly different ways. Was the core objective of trying to purchase as many shares as possible whilst appropriately managing the price risk achieved for the "Value" team? What were the total costs and how did they compare with the expected costs for the "Excess Capital" team? Look at what the execution strategy looked to maximise/minimise, were these achieved and what were their effects? This leads nicely to our next point, which is to look,through the lens of benchmarks at some of the current practices used in the market to execute share buy-backs.

**Benchmarking in Execution.** Frequently benchmarks are used as part of the evaluation process for execution quality. There is a lot of existing research and papers on execution quality, benchmarks and transaction cost analysis, which we do not wish to expand on in this paper, however there are a few points to highlight as they are applied to the execution of share buy-backs.

The first point is that when picking a benchmark to assess execution quality against, it is very important to pick one that actually helps you evaluate the performance against the Company's objectives. In this paper the "Value" camp's objectives are to buy as many shares as possible whilst taking appropriate risks. The "Excess Capital" camp's objective is to minimise value loss through the process of the buy-back execution, including the effect of any potential share price risk caused by the information lag between when the corporate actually bought the shares and the time when the shareholder has enough information to know how many shares they should sell.

The second point is understanding how much control the executing party has over the variables of that benchmark and how that affects any outcomes, especially if the broker's compensation or profit is derived in some way from the execution performance relative to this benchmark. At this point we should state that we fully expect parties to optimise any trading strategies to try to maximise the benefits to both themselves and the objectives that they are set. This paper is not trying to suggest that service providers should charge less or take risk on behalf of others for no or low compensation. We are, in part, making a point that the shareholders' interests need to be examined against fair and appropriate benchmarks.

One such Benchmark that is commonly used across the whole execution market is VWAP[17]. When VWAP is used as a benchmark normally one of two criteria are set that the executing party has no control over. These criteria are either the wall clock time that the benchmark will be set over, such as 1 hour, one day or one week. Alternatively the number of shares that need to trade in the market[18] as a percentage of the size of the order to be executed, for example if an order is 100,000 shares, and the set percent is 10%, the VWAP will be calculated from the prices of the next 1 million shares that trade in the market. Let us call this benchmark "Institutional" VWAP.

---

[17] VWAP (Volume Weighted Average Price) is a financial metric that calculates the average price of a security traded throughout the day, weighted by volume. It is often used as a benchmark for trade execution.

[18] The equity "market" is typically made up of a lot of liquidity venues and exchanges in most jurisdictions, so the exact venues and trade types that constitute the "markets volume" are pre defined between the parties involved in the trade.

# 3. Market Practices for Share Buy-Back Executions

## 3.1 Bogus Benchmark

Share buy-backs are frequently executed using execution strategies that reference a benchmark that we have (very unflatteringly) nicknamed the "Bogus Benchmark". The "Bogus Benchmark" is the arithmetic average of the daily VWAPs over the execution period.

There are several reasons why we think this benchmark is not an appropriate benchmark to use for share buy-backs, however before we go into some of those details, it is important to highlight how this benchmark differs from the very robust and well used "Institutional" VWAP benchmark.

Firstly to calculate "Institutional" VWAP over multiple days, you take the daily VWAP of each day, and weight the overall calculation by each day's volume, the "Bogus Benchmark" uses the simple average of each day. Secondly the broker has some degree of control over the time frame of the benchmark setting period, unlike the "Institutional" one, where the time frame is either set by the client beforehand, or by the market's volume. These differences can have a material effect as we will try to show later.

There are two significant factors that are relevant at this point. The first is that the "Bogus Benchmark" does not actually correlate with the objectives we stated above for either the "Value" or the "Excess Capital" camp. By this we mean that beating this benchmark does not equate directly to buying more shares or lowering the frictional cost of the capital transfer. **Secondly, this means it is not an appropriate benchmark for the company to use because, to maximise performance against this benchmark does not maximise the benefit to the shareholder. We will try to show this in an example later.**

### 3.1.1 Conflicts of Interests: A Deep Dive

Multiple conflicts of interest arise within buy-back structures based on the "Bogus Benchmark." While we won't cover all of these conflicts here, we will look at three of them at this point: the broker's compensation, the shareholders' price risks, and the guaranteed completion of the programs.

**1. Broker Compensation**

As discussed earlier, the shareholders' interests and the broker's interests are not aligned by measuring performance against this benchmark. In scenarios where the broker is compensated by outperforming the "Bogus Benchmark," there is a conflict. If the broker is guaranteeing the company a buy-back price relative to this benchmark, then the broker is exposed to risk against a benchmark. However they also wield some degree of control over how the benchmark is set, in terms of the number of trading days. As such, the broker is incentivised to optimise for their best outcome, which can have a direct conflict with the best outcome for shareholders, see example 2 below. Whilst we do not have any issue with brokers trying to maximise for their own profitability in a given transaction, this only holds true under certain circumstances which must include the product being suitable for the objectives of the corporate. In the case of a share buy-back, in our opinion it is very clear to all parties that the corporate has the fiduciary responsibility to their shareholders, so any product sold for the purpose of facilitating a share buy-back needs to uphold that responsibility.

**2. Share Price Risk**

The second conflict arises in managing the shareholders' risk concerning the fluctuation of the share price. Ordinarily, in the absence of the company's view on a share price path, a shareholder's optimal strategy is to effectively minimise this share price risk. Yet, when contrasted against strategies that are measured against the "Bogus Benchmark," the broker's interests might strongly diverge. There are instances when the broker is incentivised to keep as much optionality as possible early in the transaction, which can be achieved by trading a relatively modest proportion of the permissible daily value. This is because the broker can better manage some variables later in the transaction, and so optimise for their own interests. Refer to example 1 below. This approach is in direct conflict with the shareholders' price risk, which usually peaks early in the program. By choosing to trade less of the daily value early in the program, the broker causes the shareholder to bear a greater share price risk for a longer period, allowing the broker to have more optionality later on[19]. As previously discussed, this only seems beneficial to the shareholder if the broker or company anticipates that the share price will decline later in the program.

**3. Guaranteed Completion**

The final point on conflicts revolves around the structure of these products. In order to provide a relative price guarantee, brokers also need to offer a completion guarantee. However, it's only in the shareholders' best interest to complete a share buy-back if the purchase price/quantity of shares bought aligns with valuation considerations as discussed earlier.

Products sold with a price cap or collar embedded in their structure are relatively more expensive, as these restrictions reduce some of the broker's optionality and add hedging costs. Owing to the private nature of these contracts, hard data on this aspect is elusive. Nonetheless, research indicates that 68%[20] of Accelerated Share Repurchases (ASR's)[21], are sold without price caps or collars.

The conflict arises not in the fact that the products are more expensive, but in the fact that brokers know that if the optics of the structure do not look attractive enough to the corporate, then they are less likely to be sold. The cost of a cap depends on a set of variables such as the strike price, the stock's volatility and other factors. This means that the cost can actually mean that the corporation would have to pay a premium, rather than receive a discount to the benchmark. Asking a corporation to pay a premium would likely reduce the chance of selling the product.
.
This fact underscores the dearth of transparency and scrutiny on these products by boards, regulators, and shareholders. Understanding these details is crucial to prevent a situation where a shareholder could be forced to complete a share buy-back at a theoretically infinitely high share price, a scenario that goes against the principles of capital allocation discipline. To see an example, look at the last paragraph of Appendix 2.

**3.1.2 Reconsidering the "Bogus Benchmark"**

This subsection discusses how the common industry practice of using the "Bogus Benchmark" can blur the objectives of share buy-back execution. This practice may inadvertently lead corporations to enable an unintended transfer of value from shareholders to brokers. A critical factor lies in the seemingly innocuous choice of allowing the broker to decide the end of the trading period, a decision that could drastically shift the odds of outperforming the benchmark in the broker's favour.

We've so far approached our argument regarding the objectives of the execution phase in a binary manner. These objectives should either focus on maximising the number of shares purchased ("value") or on minimising friction ("excess capital"). However, using the "Bogus Benchmark" causes a problem by conflating these two very different goals making the outcome less clear. For instance, if you're in the "excess capital" camp, theoretically, your concern isn't about the prices at which the value is transferred, but rather that maximum value is transferred. The issue arises because most people also think that it would also be more beneficial if this transfer of value occurs at a purchase price below the average trading price over the execution period.

---

[19] For an interactive visualisation, please download and open this html file. By clicking on the 30, 60, 90, 120 day buttons you can see what happens to the shareholders residual VaR as a result of taking longer to unwind the notional value of the exposure.

[20] Chen, Kai, Press release management around accelerated share repurchases (April 7, 2020). European Accounting Review (2021) 30(1): 197–222, Available at SSRN: https://ssrn.com/abstract=3572543

[21] ASR's are a specific type of share buy-back product used mostly in the US. They are used to execute about 10% by value of S&P 500 share buy-backs.

We frequently hear this counter-argument in favour of the "Bogus Benchmark" presented as: "What is wrong with beating an average share price? Wouldn't the shareholders be thrilled if the shares were bought below the average share price over the execution period? What more could they reasonably expect?"

This is a valid question. In response, we clarify that we do not oppose an execution strategy that aims to target or beat a simple average of the share price over time, especially if the primary objective is to minimise friction. And yes, shareholders should indeed be delighted to beat this average, under one condition...

Just as most magic tricks involve a certain degree of artful deception, the act of surpassing this 'average' share price, too, is a bit like magic. 'Magic', in this context, encapsulates many nuanced layers. Herein, we're focusing on just one layer of this 'magic': the number of days in the execution period.
A Company, agreeing to an 'out-performance' fee for a broker is justified only if this broker's acumen, or their skillful application of risk, enables the company to attain superior outcomes than what could be expected without the broker's intervention. When deciding on such out-performance fees, the Company, acting in its fiduciary capacity, must reasonably anticipate that the probability of a more favourable outcome will increase due to the broker being incentivised to leverage their expertise. If this condition is met, we believe sharing some of the surplus gains accrued through the broker's efforts is wholly justified, and we generally support this practice.

To better illustrate the issue with the "Bogus Benchmark" and the implementation of these incentive fee structures, let's consider an analogy. We think that through this analogy we might provide a clearer understanding of the deception at play.

Imagine a game of coin toss where a coin is flipped 100 times consecutively. You're required to place your bet before the commencement of the game, wagering on whether there will be more heads or tails after the coin is tossed 100 times. The probability of emerging victorious in this game is unequivocally 50/50 (assuming a fair game, that is, no loaded coin etc.).

It would seem rational to pay your broker a fee if they could demonstrate, with statistically relevant data, their ability to predict the correct outcome more frequently than not, thereby enhancing your odds of winning to perhaps 55%. The additional winnings could be divided, or the broker might assure you of a 51% or 52% result, for instance.

Now imagine that the game is changed so that the coin will no longer be tossed precisely 100 times. **Instead, we can flip it anywhere from 100 to 150 times**[22]. **Here's the catch - the broker now holds the power to decide when the game ends, at any time within this range. This subtle modification of the game's rules presents a significant shift in advantage towards the broker. They can strategically halt the game when the outcome is in their favour. In light of this shift, isn't it justifiable to reassess how we incentivise the broker? After all, we're not playing the same game anymore. The broker's advantages are stacking up, and the rewards they receive should be reconsidered in line with these changes.**

Our belief is that the majority of Companies fail to appreciate this nuanced yet crucial difference between beating the average share price over the duration of the transaction and beating the average share price over a fixed timeline. In this context, the question arises: how are those who are entrusted with safeguarding the interests of the shareholders exercising their responsibilities?

The issue at hand can be approached in a multitude of ways. Brokers will often argue that in a global marketplace, competitive forces vie to deliver shareholder value. However, we cast doubt on whether this competition genuinely exists. More importantly, we question if the company or their advisors truly comprehend the revised odds of "winning this adjusted game" before agreeing to these incentive fee structures[23]. Is the new winning probability 51% or as high as 99%?

---

[22] A special thanks to "Rizzo" for helping develop the analogy.

[23] See Appendix 4 for Broker Fee Calculations.

Consider a scenario where a company hastily concurs with a figure that seems reasonable because it exceeds the 50% threshold—let's say 53%. But what if the actual winning probability is later revealed to be 95%? In such a case, the company could have unknowingly compromised a substantial portion of the shareholder's value.

The crux of the "deception" lies in the significant shift in probabilities in favour of beating this "Bogus Benchmark" if the broker is allowed to determine the end of the trading period within a pre-agreed range. In fact, as we will explore further in the examples section, the probability of falling short of this benchmark, when adopting a simple execution strategy, is virtually nonexistent. Consequently, there's a significant amount of "hidden" value that may not be immediately apparent to those without this insight.

### 3.1.3 Market Share of Buy-Back Execution Products that use the "Bogus Benchmark"

In 2022, corporate entities across the U.S. and Developed Europe bought back roughly $1.4 trillion of their own shares[24]. Over the last five years, this figure climbs to approximately $5.6 trillion. Although precise data isn't available for the percentage of these share buy-backs executed using products that are based on the "Bogus Benchmark" for their execution strategy, we do have some information about one particular subsection: Accelerated Share Repurchases (ASRs). ASRs comprise about 10% of the total value of all share buy-backs in the S&P 500.

Broker research suggests that Open Market Repurchases (OMRs) are the most prominent execution method by a considerable margin, with ASRs the second most popular execution product used by corporations. However, it's crucial to note that not all OMRs utilise the "Bogus Benchmark". To maintain a conservative estimate, let's assume that another 10% of the total value is attributed to this portion of OMRs. Using this assumption about 20% of all share buy-backs are executed using strategies with a guaranteed execution price based on an inappropriate benchmark. Assuming an average ASR buy-back contract duration of 125 days[25] and an average annualised stock volatility of 35%, we find that in 2022, at the initiation of these buy-backs, there was approximately $70 billion worth of shareholder Value at Risk (VaR)[26] managed improperly, culminating in a staggering $276 billion over the last five years. Please note that we believe this is a conservative estimation.

## 3.2 Examples of Share Buy-Backs during the Execution Phase

Let's look at two UK examples, we have anonymised the companies to protect the corporate identities. Both of these share buy-backs look like they were executed using a guaranteed product based on the "Bogus Benchmark". Of course we do not know the exact contracted detail agreed between the broker and the corporate, as they are private documents. However because of the UK disclosure rules about the daily trade details it is possible to back out certain details. We picked these two examples to show that both the "price risk" and the frictional costs were clearly not optimised for the shareholders interests. Charlie Munger, no doubt, would have predicted these outcomes if he knew the incentive structure. The first example we picked because the share price fell quite sharply, and the second one because conversely the share price rose quite sharply. We wanted to show that in both these scenarios the frictional costs of the execution contract came at a very high cost to the shareholder. It is also worth pointing out here that there is nothing clever in the actual execution strategies themselves. The core of the strategy is that the broker buys more when the stock price is trading below the rolling "Bogus Benchmark" and less when the share price is above. Anyone with some degree of technical ability can design and optimise a similar strategy, indeed you can ask a broker to do so for a cost of somewhere around 5 to 8bps (excluding advisory or banker fee if one is charged). We would imagine that shareholders would value greater disclosure in the annual reports of buy-back execution strategies employed, the risks and fees paid to do so, to prevent these sorts of outcomes repeating themselves.

---

[24] US share buy-backs $1.26tn (Bloomberg), Europe €161bn (Exane), Europe €135bn (Goldman Sachs), UK £47bn (Goldman Sachs)

[25] Chen, Kai, Press release management around accelerated share repurchases (April 7, 2020). European Accounting Review (2021) 30(1): 197–222, Available at SSRN: https://ssrn.com/abstract=3572543

[26] VaR calculation at the 32% level, i.e. using an expected price move of confidence interval of 1 std deviation (i.e. 68% confidence) = $1.4tn*20%*1*(0.35/SqRoot(250/125)) = $70bn. If using a VaR at the 5% level (corresponding to a 2 std move), this goes to $140bn.

**3.2.1 Example 1: 8.6% fee -** £15.8 million costs for a share buy-back of £184 million

In 2022, a UK FTSE 250 company embarked on a share buy-back initiative. Their goal was clear: to return £200 million to shareholders within the following nine months. The success of this endeavour hinged largely on minimising frictional costs during the execution phase.
It turned out that the total gross value of the shares purchased from the market, before tax and fees, amounted to about £184m. So, where did the remaining £16.8 go?

In the company's next annual report they reported that, including costs, the share buy-back returned £200.8m to shareholders. We examined the company's daily regulatory filings and found that the number of shares that the company purchased matched exactly with what they reported in their annual report. From the daily regulatory filings we can calculate that the shares were purchased from the market (pre tax, pre fee/commission) for a total gross value of approx £184m. The costs for this transaction appears to be £16.6m. We know that UK stamp duty is 0.5%, so by deducting this we can imply that the fee paid was approximately £15.7m, or 8.5%.

It is clear from looking at the trading pattern, that this fee was derived from purchasing a product based on the "Bogus Benchmark". Although the company has not confirmed this with us, they have told us that they knowingly entered into a risk and reward arrangement with their advisor, and that they know it ended up costing them a lot of money. We are less interested in what it costs per se, we are more interested in the structure itself, and is it designed to be one that attempts to maximise for the shareholders best interests? Buying risk protection against an event that did not happen might well still have been in the best interest of the client. If that was not true then what would happen to our insurance industry?

We think there are really three questions to answer- the first is what risk is the product that they bought designed to protect against? The second : is this actually a risk aligning with lowering the frictional cost to the shareholder? The third: is the price that they paid fair? So let's try to answer them, using the imperfect information we have available to us.

1. The product protects against the company's average purchase price for the share buy-back being above the "Bogus Benchmark". We assume that the structure was probably what is nicknamed a "VWAP - Minus[27]" product ("the structure"). So lets say that the broker guaranteed that the client would buy their shares at "VWAP" minus say 0.5%.
2. Does the "structure" align with minimising or managing the frictional cost of transferring £200m from the company to the shareholders via buying shares? No it does not.
3. Is the price they paid for the risk protection fair? Let's not answer this question because the answer to question 2 makes this a fairly mute point. Having said that, the actual "price" they paid is not possible to calculate without a lot more information. The reason for this is because the value of the "structure" to the broker is based on its "option" value. What the company is actually doing through the structure is selling the "value" of their share price volatility for some guaranteed "out performance". To price this you need to know what is the probability of this structure yielding a value that is higher than say the "0.5%" out performance that we guessed at in answer 1. What we do know however is that a very basic execution strategy of buying more value when the share price is below the "benchmark" and less when it above makes it almost impossible to underperform this benchmark. In order to underperform you would need the stocks volatility to decline to close to zero over the trading window, which is highly unlikely. The scenario where that might happen is if the company was subjected to a take over bid, however we understand that it is normal for brokers to negate this scenario in their contract documents.

---

[27] "VWAP-Minus", "Optimised-VWAP", "Guaranteed Sub-VWAP", "VWAP-Guaranteed" and a variety of other names are given to some of the products sold to corporations for the execution of buy-back. See Appendix 4 for a bit more detail.

The simple reason we answered question 2 with a no is because the frictional costs of the transaction are directly related to the size of the "out performance". For the structure to make sense from the perspective of meeting the shareholder objectives then the friction would need to reduce as the out performance increased. This is not how it is designed. Rather than us trying to explain this in more detail, it is easier to just show this to you in Example 2.

We question if the company understood the nuances of the structure that they bought, both from a shareholders risk perspective, and the cost perspective? The consequence of this decision on the cost alone, is that they appear to have paid a fee that was 100 times higher than the equivalent agency fee. We conclude that unfortunately the company unwittingly gave away a tremendous amount of shareholder value, for a virtually worthless perceived benefit, while the share price risk was not managed appropriately. If this product was sold as a "Win-Win" product, then truly there really only was one winner, the broker.

In the observed share buy-back program in Figure 1, the company achieves 100% volume completion (i.e., all designated shares for the buy-back have been purchased) in only 39% of the allowable time for the share buy-back. This means the company exhausted its buy-back allocation much earlier than the end of the stipulated time period.

The early completion of the buy-back could potentially mean that the company might be missing out on potentially more favourable pricing opportunities in the later stages of the allowed time.

**Figure 1**. Timing Mismatch: Volume Completion Versus Allowed Buy-Back Period

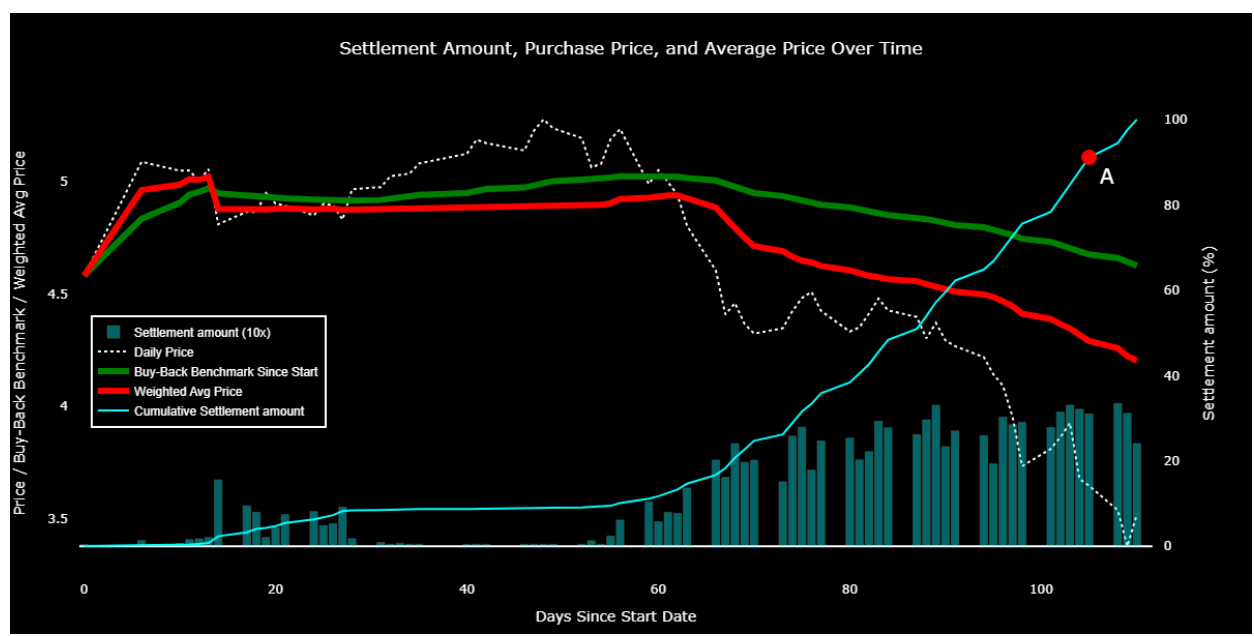

### 3.2.2 Example 2 - A share buy-back of £435m at a cost of £10m

In the first half of 2023, a UK FTSE 100 company completed a share buy-back program. Their objective was transparent and strategic: to return a hefty £435 million of 'surplus capital' to their shareholders within the following six months. Their course of action, however, was characterised by the purchase of approximately 90% of the value of shares in the first three months, and stretching out the acquisition of the remaining 10% over the same time-frame again, another three months.The final outcome evidenced a prioritisation of outperforming the set benchmark rather than optimising for the shareholders' best interests, leading to a performance fee approximately 30 times higher than the equivalent agency fee. Consequently, it's evident that while the company's initial intent was to efficiently return surplus capital to shareholders, the execution strategy fell short in managing the frictional fees effectively.

We delve into this in more detail: In the announcement of the buy-back, the reason given was to return surplus capital to shareholders. The 2022 annual report was released after the execution of the buy-back had completed. We applaud the companies transparency in their reporting, they clearly noted that they paid £10m in performance fee (2.2%).

If we examine the company's daily regulatory filings and recreate the trading pattern, we see that the company purchased approximately 90% of the value of the shares in the first three months of the program whilst the share price was relatively weak. They then took the remainder of the allowed period to purchase the residual 10% of the value as the stock price rallied to its peak price in the last weeks of the program.

The question we are trying to answer in this example is: what effect does the broker optimising for "out performance" have on the frictional cost to the shareholder.

At point A on in Figure 2, 90% of the value has been executed. The average purchase price is ~1% below the benchmark price at the point.

This second example shows that optimising for out performance is independent of the share price being above or below the benchmark, and through slowing the pace of the execution right down when the share price is above the "Bogus Benchmark", the broker has managed to increase the "out performance" by 265%, while the average purchase price also increased by 0.6%.

The exact contractual details of how the performance fee is calculated from the out performance is not disclosed, but it is very clear what the execution strategy is optimising for. Note that in this example the performance fee is paid directly by the corporate. So although it is not paid for by the portion of the company's capital allocated to the share buy-back as it is in Example 1, it is still the company's capital which foots the bill, this however is still 100% shareholder's capital.

We believe this example shows that the broker involved was responding to the incentive of outperforming this benchmark, rather than optimising for the best outcome for shareholders. We asked the company to explain the nature of the performance fee and they confirmed that they bought a product from their broker that they described as an "enhanced agency model that guaranteed a discount to the VWAP". We checked if the VWAP was what we call "Institutional" VWAP, however they confirmed it was what we call the "Bogus Benchmark".

Again we question if the company understood the nuances of the structure that they bought, both from a shareholders risk perspective, or cost perspective? It means that they bought price protection against a benchmark that was not relevant to their objectives. The consequence of this decision on the cost alone, is that they appear to have paid a fee that was approximately 30 times higher than the equivalent agency fee[28]. Given that the language in the original buy-back announcement says the reason for the buy-back was "to return surplus capital to shareholders" it seems fair to put the reason for this buy-back in the "excess capital" camp and conclude that the frictional fees were not well managed.

**Figure 2**. Trade Volume and Price Dynamics during a Share Buy-back with a front-loaded profile

---

[28] It is not normally sensible to compare agency commission with a principal risk fee, they are like apples and oranges. However in both Example 1 & 2, all of the shares bought by the broker, are sold to the corporate at exactly the same price and on the same day that they were bought on. This is detailed in both of the regulatory notices published by the companies when they announced the buy-backs. Since the actual executions are agency-like, we feel that this comparison is fair. It is the contractual details agreed as they relate to the "Bogus Benchmark" that explains the fees.

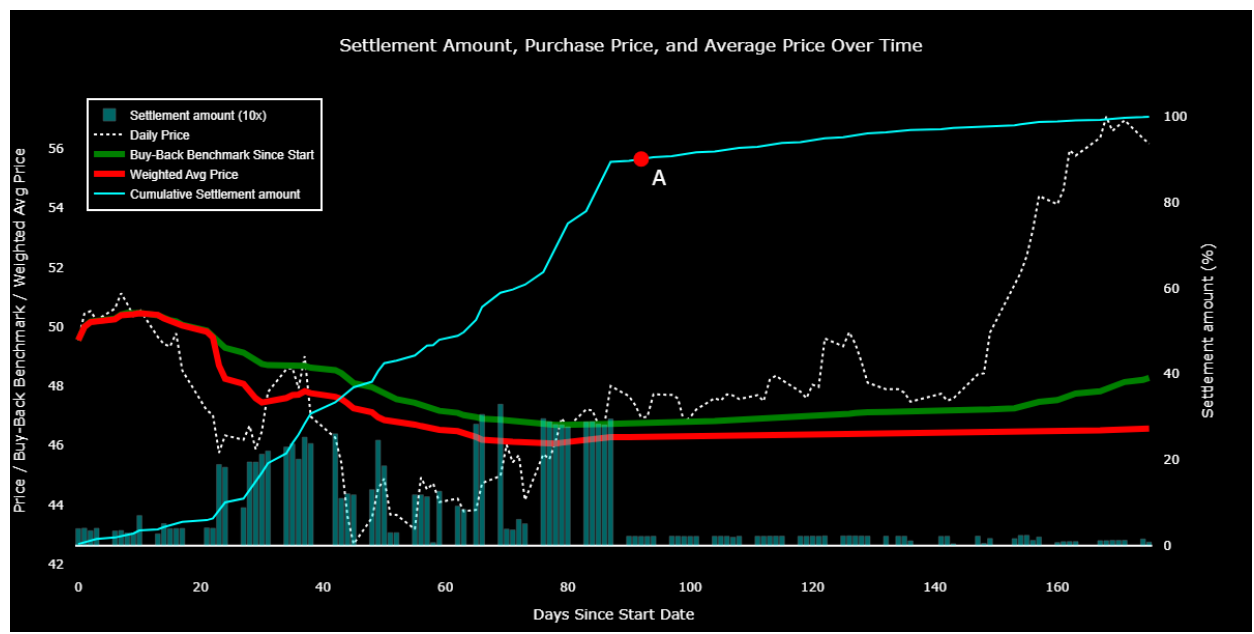

### 3.2.3 Comparison of Example 1 and Example 2

We delve deeper into the intricacies of the share buy-back execution process, by comparing key characteristics of our two examples.

Table 1 outlines the key differences in the execution processes for two different cases, Example 1 and Example 2, at the stage where 90% of the share buy-back value has been completed (Point A). This comparison aims to offer insight into the drivers behind the broker's decision-making and the ensuing impacts on the shareholders.
In the table, various performance metrics are compared, including the percentage of value executed, percentage of allowable time expired, and the percentage outperformance. It also compares the share price to the benchmark and the influence of a 1-day use of the previous day's share price on the benchmark, as well as the effect of 1% value (using the previous day's price) on the average share price, and finally, the effect of a 1% value on performance using the previous day's share price.

**Table 1**: Comparison of key characteristics of the execution process for Example 1 and 2

| Point A (90% completion of the share buy-back) | Example 1 | Example 2 |
| --- | --- | --- |
| % value executed | 89.3 | 89.7 |
| % allowable time expired | 37.5 | 53.0 |
| % out performance | 8.2 | 1.0 |
| Share price to Benchmark price (at Point A) | -22.0% | 2.0% |
| Effect of 1 Day using prev. day share price on benchmark | 0.31% | -0.03% |
| 1% value using prev. day price on average share price | 0.19% | -0.03% |
| Effect of 1% Value on Performance using prev day share price | 2.13% | -3.19% |

Both trades are approximately equally completed, with 90% of the value of the buy-back already executed. Both programmes still have lots of time left, so the brokers are under no time pressure to complete quickly.

Embedded in this available time left is value, but it is hard to see this value optically. Think of it as the longer you have left in the programme the greater the possibility of the share price moving in that time frame, or "theta" if you are options literate. That is pretty much where the similarity ends between the two examples.

The 90% of value of the executed portion in Example 1 is currently 8.3% below (out-performing) the "Bogus Benchmark", and the current share price is a whopping 22% below the benchmark. Whereas the 90% of value in Example 2 is currently only 1% below (out-performing) the benchmark, and the current share price is now 2% above this benchmark.

Let's assume that neither broker has a crystal ball, so they do not know or have a high confidence in what the share price is now likely to do over the rest of the allowable trading window. How should each broker think about the way to create the maximum out performance from here? Let's see how each of these decisions affects the shareholders.

In Example 1, if we roll the clock forward one day, and keep all the other factors the same, because the share price is 22% lower than the benchmark, when you add that extra day's price to the average it moves the benchmark ⅓ of a percentage point lower. The 8.3% outperformance goes to 8%, roll it forward 3 days a full percent is lost in performance. This means that if the broker wants to try to actually increase the out performance they need to trade faster than 1.5% of the total value each day, just to keep the out performance from dropping. So they do that and the trade is completed in 3 more days, and they increased the out performance to about 9.1%. The remaining time value in the trade was "wasted", but the reward was pretty spectacular all the same.

The net effect of the broker speeding up as the share price falls, was to grow the out performance. If you roll the clock on further you can see that the shareholder would have actually been much better off trading slower as there was so much time left and the stock price ended up ever lower still, but that is a little unfair on the broker since we have already said that they did not have a crystal ball. The real problem is that almost none of that benefit of buying the "cheap shares" went to the shareholders as it was pretty much all paid out in fee. The broker bought the shares at a 9.1% discount to the average share price over that time, but the shareholder, after fees, bought them at a price that was ~8.5% higher than where the broker bought them, and this is before paying the tax.

The situation in Example 2 is actually more interesting, even if it is less favourable for the broker. Since the stock price is now currently above the benchmark, the game for the broker is to slow down, unless the share price falls below the benchmark. The broker has a lot of time left, with nearly 50% of the days still available to use. For every day that the stock price is above the benchmark, so long as the broker buys very few shares each day, then the performance will widen. This is because the average of the shares bought so far is not really moving much higher (because the broker is buying only a few shares each day), but the benchmark is ticking higher at a much faster rate with each day. Clearly the broker does not know what the share price path is going to do, but so long they do not buy more value, as percent of the total value, on any given day, than that day's contribution to the benchmark the performance will widen. In this case the total available trading window was about 120 days, so each day's VWAP price was a little less than 1% of weight of total. But they had about 60 days left to buy 10% of the value, or about 0.15% of the total value each day. This means they had time on their side. If the share price had dropped below the benchmark they would have probably sped up and completed, but it did not, so they traded less and less each day and the out performance increased to about 3.5% over the second half of the programme. At the same time the average share price got slightly higher, and therefore the shareholder not only paid a higher price for their shares, but also a higher fee due to the out performance increase. It is not hard to see that this increasing out performance did not benefit the shareholder in any way, in fact their average share price went slightly higher than when the programme was 90% complete. But the portion of the pie that incentivised the broker increased from ~1% to ~3.5%, or from £4.3mil to £15.1mil.

What we have shown in both these examples is that the broker can increase the out performance when the stock price is either above or below the benchmark price. By incentivising them to increase the "out performance" bucket, all we are doing is increasing the friction paid by the company with no benefit to the shareholder, in fact in Example 2 you can argue that there is also a direct cost to the shareholder if they choose to hold the shares longer term, as the average price of the shares increased due to the strategy, which means less shares were bought in total.

# 4. Conclusion

Company boards and executive management teams must protect shareholders' interests by ensuring that the objectives of the share buy-back get translated into an appropriate execution strategy. The two main objectives that boards normally state for share buy-backs are either to reduce share capital or to return excess capital.

We have found that certain buy-back execution products reference "VWAP" and use incentive fee/profit structures that escalate when outperforming this "VWAP". We call this "VWAP" the "Bogus Benchmark" to differentiate it from traditional VWAP. These products do not optimise for either of the two main objectives we mentioned above, and therefore, are not fit for purpose. **In our opinion, these products are designed to ensure[29] that the broker can sell the shares to the corporate at a PREMIUM to the price at which they were bought in the market.**

**Why should we care?** Consider this - over the past 11 years, a minimum of 20% of Apple Inc's share buy-backs were executed using execution products that reference this "Bogus Benchmark". The unwarranted fees and profits extracted through the use of these products have resulted in staggering shareholder losses. In Apple's case we estimate these fees/profits exceed $550mil. Had this money been used to buy more shares, as was intended, then Apple shareholders would have gained an extra $6 billion in capital gains alone. This isn't just an Apple issue; it's been happening across the market for at least 25 years.

When discussing the execution phase of share buy-backs, there are still several issues that need to be addressed. While we intend to cover some of these issues in later work, we first seek feedback, to listen, and engage in discussion, and to learn and correct our mistakes. In later work, we plan to delve deeper into the subject of ASR's, the execution expertise of corporate advisors, the new SEC disclosure rules, and relevant market structure topics. These include anti-gaming approaches and enabling greater use of all available liquidity venues to allow corporations to source liquidity comparable to how institutions do.

Our companies require trustworthy guidance and independence when purchasing equity execution products especially when these products have designed conflicts. Failure to provide appropriate products at reasonable fees will erode their capital, diminish their confidence, and ultimately harm our capital markets. It is imperative that companies prioritise the interests of their shareholders and seek trustworthy guidance to safeguard their investors' capital and confidence in the market.

---

[29] There is a slight risk to the broker that they might owe a portion of the discount offered if the volatility of the stock falls significantly during the execution phase. However this discount is to a benchmark that has no relevance to the shareholders best interests in the first place.

# Appendix 1 - Risk Deep Dive

## The Stakes at the Start of the Implementation Phase

At the inception of a share buy-back program, the scale of the commitment made by a corporation is substantially quantified by the total value of the planned share repurchase. Although this commitment remains constant, the number of shares that can be purchased with this static value fluctuates based on the dynamics of the share price during the execution period.

For a more concrete understanding, let's shift our focus to 2022. In that year, the total volume of share buy-backs reached $1.4 trillion. We conservatively estimate that at least 20% of this sum, amounting to $280 billion, was executed via share buy-backs using execution strategies that are designed to out-perform the "Bogus Benchmark", and hence meaning the share price risk has been mis-managed. This scenario provides us with an illustration of the immense financial stakes involved at the onset of the execution phase.

Let us first look at single buy-back, and then extrapolate that logic over the whole market for 2022 using the same assumptions, to try to give an idea of the scale of the issue as we see it. Single buy-back: Assume that, based on historical trends, the company's share price has a volatility level of 35% per year and the execution of the buy-back program is expected to take six months (According to Chen (2021)[30], the average ASR characteristics indicate a mean contract period of 125 days.).

The VaR at a 95% confidence level could be estimated using standard methods. The result will represent the minimum loss expected (with a 5% likelihood), given the size of the buy-back and the time required to execute it, due to adverse price movements.

If we denote the buy-back value as V, the stock's return volatility as σ (standard deviation of the stock's return), the length of the execution period as T (measured in years), and the z-value corresponding to a specific confidence level (say 95%) as z, then the VaR for the buy-back 'portfolio' at the beginning of the execution period can be estimated as:

$$VaR = V * z * \sigma * \sqrt{T}$$

The VaR measure hence provides an estimation of the value at risk due to potential adverse price movements during the execution phase of the buy-back program. An increase in the stock's return volatility (σ), the length of the execution period (T), or both would lead to a higher VaR, indicating a higher potential loss.

Figure 3 presents a Monte Carlo simulation of the potential risk exposure based on the total share buy-back volume over the execution horizon. The X-axis represents the progression of the execution, while the Y-axis denotes the stock price movement.

The simulation is shown as a series of data points representing the predicted stock price at each point in time. Additionally, two curves are highlighted on the chart: one representing the 1% VaR and the other showing the 5% VaR.

These curves indicate the maximum expected loss (with 1% and 5% likelihood, respectively) due to adverse price movements over the execution horizon. They are plotted as positive numbers to reflect the potential loss or 'risk exposure' associated with the execution of the share buy-back.

---

[30] Chen, Kai, Press release management around accelerated share repurchases (April 7, 2020). European Accounting Review (2021) 30(1): 197–222, Available at SSRN: https://ssrn.com/abstract=3572543

**Figure 3.** Risk Exposure in Share Buy-backs: A Monte Carlo Simulation

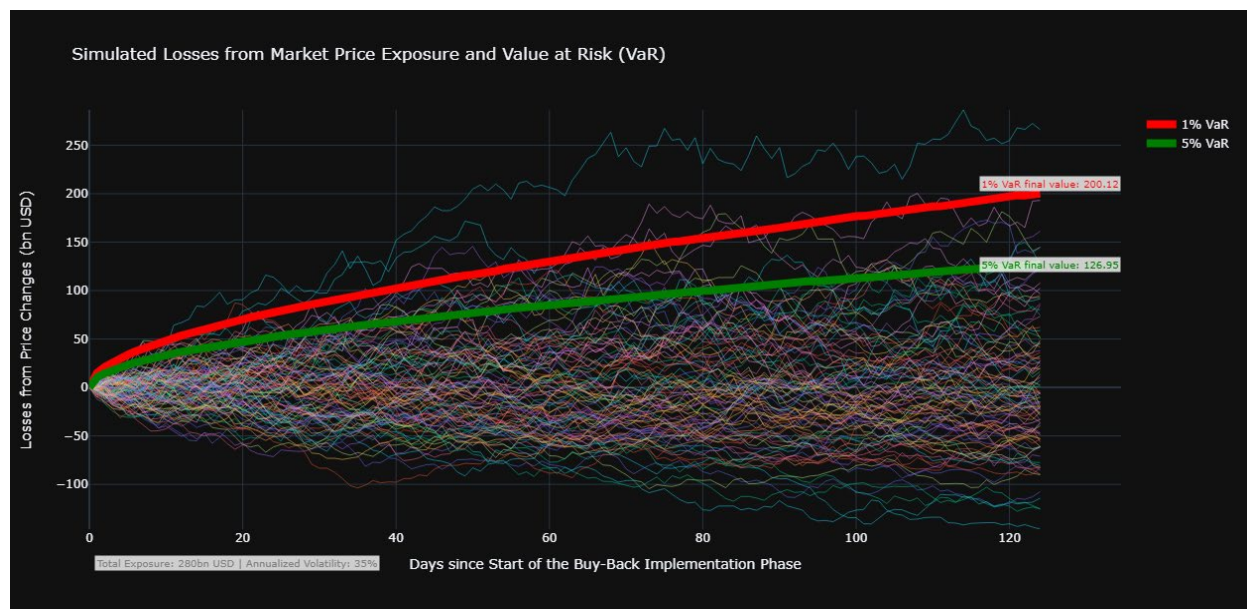

Now let's apply that logic to the whole market, to put some values on the scale of the issue, using the same assumptions of volatility and length of buy-back program we used in our single buy-back.

Here we have employed a Monte-Carlo Simulation to calculate Value at Risk (VaR), using geometric Brownian motion as our model. Following the popular approximation for VaR, we can also calculate the total risk exposure for 2022: $280bn * 2.33 * 0.35 * sqrt(252/125) = $161bn for the 99% VaR and $280bn * 1.96 * 0.35 * sqrt(252/125) = $135bn for the 95% VaR. These results are closely aligned with the Monte-Carlo simulated numbers illustrated in Figure 3.

Considering the entire span of the past five years, the risk exposure escalates to $643 bn (99% VaR) and $541 bn (95% VaR).

## Timing of Execution: Extent of Exposure

The timing of execution, or the time duration over which a buy-back is spread, plays a crucial role in determining a buy-backs exposure to unpredictable market prices. The longer the length of execution time, the longer the duration of exposure to these market price fluctuations. Let's examine the implications of this with the help of a trading strategy that involves trading Time Weighted Average Price (TWAP) over 120 days.

With the TWAP strategy, the company would distribute its trades evenly throughout the trading period to minimise the impact on the market price. This strategy assumes a constant trading value each day for day, and we have chosen 120 days for our example. As the execution progresses the remaining value to be bought back, and hence the exposure to future price fluctuations, gradually diminishes. This reduction in exposure over time translates into a decrease in the residual risk (defined here as Value at Risk (VaR) from the current point in time until the end for the remaining value has been traded).

However, it's important to understand that while spreading the buy-back over a more extended period decreases the daily value to be traded, and so reduces the daily impact, it also extends the duration of exposure. This means that if the market trends upward during the execution period, the company may end up buying back shares at higher prices, thereby eroding shareholder value.

A Monte Carlo simulation can effectively illustrate this trend. By running numerous scenarios and plotting the residual VaR over time, we can observe how this risk is reduced as the execution of the buy-back progresses.

**Figure 4**. Residual VaR Reduction over Time in TWAP Execution

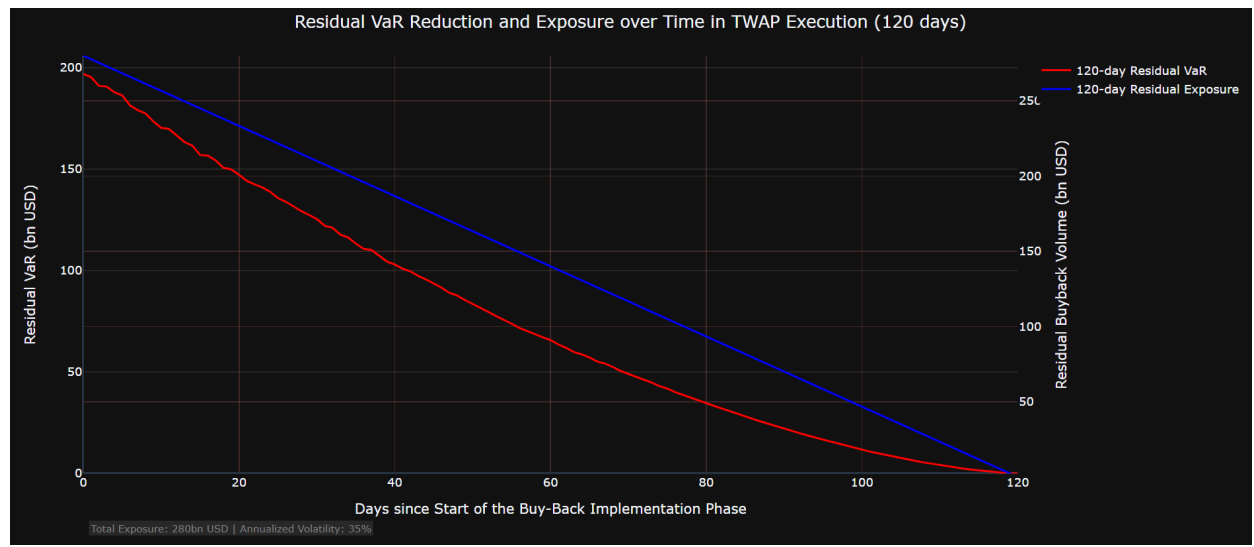

Figure 4 shows the number of days into the execution on the X-axis and the residual VaR on the Y-axis. The line on the chart starts at a high point on the Y-axis at day 0, indicating a high VaR due to the entire volume being yet to be traded. The line then gradually declines as the number of days increases, illustrating the reduction in residual risk over the execution period.

This chart illustrates how the residual VaR decreases over time during the execution of a share buy-back following the TWAP strategy over 120 days. It highlights the risk exposure diminishing over the execution period, showcasing the importance of strategic timing during the execution phase.

For an interactive visualisation, please download and open this [html file](). By clicking on the 30, 60, 90 and 120 day buttons on the top of the chart, you can see how the residual VaR changes with the duration of exposure. Even though the notional value of the risk is identical on day 0 in all 4 trading duration scenarios, the difference to the residual VaR for a risk unwind strategy that will take 120 days is approx 110% higher than one that will take 30 days to unwind.

## Appendix 2 - Investment Trust Hypothetical Example

A UK listed investment trust is a type of investment fund that is constituted as a public limited company and is listed on the London Stock Exchange. It invests in other companies with the aim of generating profit for its shareholders. Unlike other types of investment funds, investment trusts are closed-ended, which means that the fund managers cannot redeem or create shares.

Our Hypothetical example, HYPO, is an investment trust that has a single investment of 1mil shares in another listed company called INV. INV shares trade at a value of £100. HYPO has an outstanding share count of 10 mil shares, and currently trades at a share price of £7.
Lets imagine our company has no operating expenses or liabilities of any sort.
The current NAV (Net Asset Value - think of this as "Intrinsic Value") of the fund is therefore £100mil, and the market cap of the investment trust is £70mil. This means that the investment trust is currently trading at a discount of 30% to its NAV.

INV pays a 10% dividend, in cash in our perfect 0% tax environment, so the fund now has two investments, 1mil shares of INV plus £10m in cash.
Management recommended to the board that the fund should execute a share buy-back of £10m.
The company makes the public announcement that they have approved a share buy-back for £10m.

In this very simple example, it is very easy to see that if the fund could buy 14% of its outstanding shares (1.43m shares * £7 = £10m).

If for example the share price rallied to £11 on day 1 of the share buy-back, before a single share had been bought (and price of INV remained unchanged), the NAV discount has now disappeared, and the shares are trading at a 10% premium to NAV. If the Company allowed the share buy-back to progress, this £10m would only buy 9% of the outstanding shares with the share buy-back clearly eroding value for the residual shareholders.[31]

Hopefully this example helps explain why the valuation methodology needs to be understood before an execution strategy is designed and committed to.

Now let's imagine that our company bought a "guaranteed VWAP and guaranteed completion" product from a broker. If the product has no price cap, then there is no reasonable way to protect the residual shareholders from the value that is going to be destroyed if the shares are all bought at £11.

## Appendix 3 - Buy-Back Regulations

Jurisdiction-specific regulations significantly impact the execution of share buy-backs. In the US, for instance, safe harbour provisions offer legal liability protection to companies that adhere to the rules. Generally, four critical components are incorporated into regulations specific to buy-backs across most jurisdictions:

1. **Disclosure:** This mandates the communication of the company's overall intentions, such as the buy-back size, purpose, and dates, to the market.

2. **Progress Update:** This involves providing regular updates to the market on the execution progress, with reporting frequency varying from daily (UK), weekly (EU), to quarterly (US).

3. **Market Abuse:** These rules aim to limit stock price manipulation by imposing constraints on maximum trading volume, trading times, and price-related factors. For instance, regulations may restrict buy-backs from contributing to an "up-tick" in prices or exceeding a certain percentage of the trading volume.

4. **Material Nonpublic Information (MNPI)[32]:** These regulations determine the windows during which the Company can alter or initiate buy-backs.

However, these regulations have nuances that need careful consideration during a buy-back's execution. Two significant points are worth noting. Firstly, the disclosure statements primarily communicate intentions. These intentions are not set in stone and can be adjusted if market conditions change significantly during the buy-back period. For example, a substantial rally in the share price might necessitate a reassessment of the original decision, and an alternative capital allocation strategy might be more beneficial for long-term shareholders.

---

[31] This example also shows that the real value to the long-term shareholder is capturing the discount. If the value of INV was to rise 10%, and the value of HYPO was to rise 20%, then the discount has shrunk, and therefore the value of doing a buy-back has reduced. In this example building an execution strategy around the "discount to NAV" makes a lot more sense than building on the absolute price of just HYPO.

[32] Material Non-Public Information – to try to prevent/limit insider trading – since the board and management have more information about the company. E.g. safe harbour 10b5-1 in the US

Secondly, the Market Abuse regulations impose specific share volume restrictions. These restrictions dictate what percentage of a company's total daily trading volume can be dedicated to the buy-back on any given day. This volume percentage directly impacts the time it takes to execute the buy-back.

While this provides an outline, bear in mind that the specifics can vary greatly across jurisdictions.

## Appendix 4 - Understanding Broker Fees for "VWAP Guarantee" and "VWAP - Minus" products

A broker fee is a cost that a corporation has to pay the broker for managing their share buybacks. How this fee is calculated depends on the broker's performance compared to a pre-set standard, which we are referring to as the "bogus benchmark".

The specifics of this fee calculation can vary based on the agreements between the corporation and the broker. Usually, these agreements are of two types: one that guarantees a certain level of outperformance (called "VWAP Guarantee"), and another that promises a minimum outperformance with a share in any additional outperformance (called "VWAP - Minus").

**Guaranteed Product (often referred to as "VWAP Guarantee")**

The broker fee for the "VWAP Guarantee" product is the total outperformance, minus the agreed guarantee. For example, if the broker managed to outperform the bogus benchmark by 1% (100 basis points, or bps), and the guaranteed outperformance was 40bps, then the broker fee would be 100bps (total outperformance) - 40bps (guarantee) = 60bps.

So in this case, the fee calculation or "function" is: (total outperformance - guaranteed outperformance).

**Out performance Product (often referred to as "VWAP - Minus")**

For the "VWAP - Minus" product, the fee is a percentage of the amount by which the total outperformance exceeds a minimum guaranteed outperformance.

For instance, if the broker outperformed the bogus benchmark by 1% (100bps), and the minimum guaranteed outperformance was 30bps, and the broker is to receive 70% of any further outperformance, then the broker's fee would be (100bps - 30bps) * 0.7 = 49bps.

So, in this case, the shareholders retain a portion (in this example, 30%) of any outperformance beyond the minimum guaranteed level.